\documentclass[11pt,draftcls,onecolumn]{IEEEtran}

\usepackage{graphicx}
\graphicspath{{figures/}}
\usepackage{amssymb,amsmath,bm}
\usepackage{cite,url}
\newcommand\mymatrix[1]{\bm{\mathrm{#1}}}
\newtheorem{property}{Property}

\newtheorem{proposition}{Proposition}

\usepackage{color}
\definecolor{dgreen}{rgb}{0,.6,0}
\usepackage[normalem]{ulem}
\usepackage{lineno}
\newlength\halfwidth
\newlength\figwidth
\begin{document}

\title{A Differential Cryptanalysis of Yen-Chen-Wu Multimedia Cryptography System (MCS)%
\thanks{Chengqing Li was supported by The Hong Kong Polytechnic University's
Postdoctoral Fellowships Scheme under grant no. G-YX2L. Shujun Li
was supported by a fellowship from the Zukunftskolleg of the
Universit\"at Konstanz, Germany, which is part of the
``Exzellenzinitiative'' Program of the DFG (German Research
Foundation). The work of Kowk-Tung Lo was supported by the Research
Grant Council of the Hong Kong SAR Government under Project 523206
(PolyU 5232/06E).}}

\author{Chengqing Li\thanks{Chengqing Li and
Kowk-Tung Lo are with the Department of Electronic and Information
Engineering, The Hong Kong Polytechnic University, Hung Hom,
Kowloon, Hong Kong SAR, P.~R. China.}, Shujun~Li\thanks{Shujun Li is
with Fachbereich Informatik und Informationswissenschaft,
Universit\"at Konstanz, Fach M697, Universit\"atsstra{\ss}e 10,
78457 Konstanz, Germany.}, Kowk-Tung~Lo and Kyandoghere
Kyamakya\thanks{Kyandoghere Kyamakya is with Universit\"{a}t
Klagenfurt, Institut f\"{u}r Intelligente Systemtechnologien,
Universit\"{a}tsstra{\ss}e 65-67, 9020 Klagenfurt, Austria.}}

\maketitle


\begin{abstract}
At ISCAS'2005, Yen et al. presented a new chaos-based cryptosystem
for multimedia transmission named ``Multimedia Cryptography System''
(MCS). No cryptanalytic results have been reported so far. This
paper presents a differential attack to break MCS, which requires
only seven chosen plaintexts. The complexity of the attack is
$O(N)$, where $N$ is the size of plaintext. Experimental results are
also given to show the real performance of the proposed attack.
\end{abstract}

\begin{keywords}
chaos, cryptanalysis, differential attack, encryption, multimedia,
security
\end{keywords}

\section{Introduction}
\setcounter{equation}{0} \noindent

The prevalence of multimedia data makes its security become more and
more important. However, traditional cryptosystems can not protect
multimedia data efficiently due to the big differences between texts
and multimedia data, such as the bulky sizes and strong correlation
between neighboring elements of uncompressed multimedia data. In
addition, multimedia encryption schemes have some special
requirements like high bitrate and easy concatenation of
different components of the whole multimedia processing system. So,
designing special encryption schemes protecting multimedia data
becomes necessary. To meet this challenge, a great number
of multimedia encryption schemes have been proposed in the past two
decades \cite{Bourbakis:SCAN:PR1992, Chung-Chang:SCAN:PRL1998,
Josef:ChaoticImageEncryption:JEI98,
Fridrich:ChaoticImageEncryption:IJBC98, YaobinMao:CSF2004,
Wu&Kuo:MHTEncryption:IEEETMM2004, Flores:EncryptLatticeChaos06,
Pareek:ImageEncrypt:IVC2006, Xiao:ImproveTable:TCASII06,
Kim:ArithmeticCoding:TSP07, Wong:ChaosEncrypt:IEEETCASII08}. Due
to the subtle similarity between chaos and cryptography,
some of multimedia encryption schemes were designed based
on one or more chaotic systems
\cite{Josef:ChaoticImageEncryption:JEI98,
Fridrich:ChaoticImageEncryption:IJBC98, YaobinMao:CSF2004,
Pareek:ImageEncrypt:IVC2006, Xiao:ImproveTable:TCASII06,
Wong:ChaosEncrypt:IEEETCASII08}. Meanwhile, a lot of
cryptanalytic work has also been reported, showing that many
encryption schemes were not designed carefully and are prone to
various kinds of attacks \cite{Jan-Tseng:SCAN:IPL1996,
Yu-Chang:SCAN:PRL2002, Lian:BreakFridrich:PA05,
Solak:BreakObserver:IJBC05, AlvarezLi:Cryptanalysis:PLA2005,
Kaiwang:PLA2005, David:AttackingChaos08, Zhou:AnalysisHuffman2007,
Rhouma:BreakLian:PLA08, Goce:cryptanalysis:TM08,
Zhou:AttackArithmeticCoding:TSP09, LiLi:IVC2009b}.

In the past decade, a series of encryption schemes were proposed by
Yen and Guo's research group \cite{Yen-Guo:HCIE:IEEPVISP2000,
Guo-Yen-Pai:HDSP:IEEPVISP2002, Chen&Yen:RCES:JSA2003,
Chen-Guo-Huang-Yen:TDCEA:EURASIP2003, Yen-Guo:MCS:ISCAS2005}. The
main idea of these schemes is to combine some basic encryption
operations, under the control of a pseudorandom bit sequence (PRBS)
generated by iterating a chaotic system. Unfortunately, most of
Yen-Guo multimedia encryption schemes have been successfully
cryptanalyzed \cite{Li:AttackTDCEA2005,Li:AttackDSEA2006,
Li:AttackHDSP2006, Li:AttackingPOMC2008, Li:AttackingRCES2008}.

This paper reports a security analysis of MCS (Multimedia
Cryptography System) -- the latest multimedia encryption scheme
proposed by Yen et al. \cite{Yen-Guo:MCS:ISCAS2005}.
Another hardware implementation of MCS was
proposed in \cite{Yen:MCS:ISCE2007}. Compared with other earlier
designs, such as RCES \cite{Chen&Yen:RCES:JSA2003} and TDCEA
\cite{Chen-Guo-Huang-Yen:TDCEA:EURASIP2003}, which have been
cryptanalyzed in \cite{Li:AttackingRCES2008, Li:AttackTDCEA2005},
MCS combines more encryption operations of different kinds in a more
complicated manner, in the hope that the security can be effectively
enhanced. This paper shows that MCS is still vulnerable to a
differential chosen-plaintext attack. Only seven chosen plaintexts
(or six specific plaintext differentials) are enough to break MCS,
with a \textit{divide-and-conquer} (DAC) strategy.

The rest of this paper is organized as follows.
Section~\ref{sec:scheme} briefly introduces how MCS works. The
proposed differential attack is detailed in
Sec.~\ref{sec:Cryptanalysis} with experimental results. Finally the
last section concludes the paper.

\section{Multimedia Cryptography System (MCS)}
\label{sec:scheme}

MCS encrypts the plaintext block by block, and each block contains
15 bytes. As the first step of the encryption process, each 15-byte
plain-block is expanded to a 16-byte one by adding a secretly
selected byte. Then, the expanded block is encrypted with the
following four different operations: byte swapping (permutation),
value masking, horizontal and vertical bit rotations, which are all
controlled by a secret PRBS.

Denote the plaintext by $f=(f(i))_{i=0}^{N-1}$, where $f(i)$ denotes
the $i$-th plain-byte. Without loss of generality, assume that $N$
can be exactly divided by 15. Then, the plaintext has $N/15$ blocks:
$f=(f^{(15)}(k))_{k=0}^{N/15-1}$, where
$f^{(15)}(k)=(f^{(15)}(k,j))_{j=0}^{14}=(f(15k+j))_{j=0}^{14}$.
Similarly, denote the ciphertext by $f'=(f'(i))_{i=0}^{(N/15)\cdot
16-1}=(f'^{(16)}(k))_{k=0}^{N/15-1}$, where
$f'^{(16)}(k)=(f'^{(16)}(k,j))_{j=0}^{15}=(f'(16k+j))_{j=0}^{15}$
denotes the expanded cipher-block. With the above notations, MCS can
be described as follows.

\begin{itemize}
\item
\textit{The secret key} includes five integers $\alpha_1$,
$\alpha_2$, $\beta_1$, $\beta_2$, $Secret$, and a binary fraction
$x(0)$, where $1\leq\alpha_1<\alpha_1+\beta_1\leq 7$,
$1\leq\alpha_2<\alpha_2+\beta_2\leq 7$,\footnote{In
\cite{Yen-Guo:MCS:ISCAS2005} Yen et al. didn't exclude the
possibility of $\alpha_i=0$ and $\beta_i=0$, but to achieve the
effect of encryption they should not be equal to 0.}
$Secret\in\{0,\ldots,255\}$ and $x(0)=\sum_{j=-64}^{64}x(0)_j\cdot
2^j$, $x(0)_j\in\{0,1\}$.

\item
\textit{A PRBG (pseudorandom bit generator)}

A pseudorandom number sequence $(x(i))_{i=0}^{N/15-1}$ is generated
by iterating the following equation from $x(0)$:
\begin{equation}
x(i+1)=\left((419/2^8)\cdot\left(x(i)\oplus H(x(i))\right)\bmod
2^{64}\right)\bmod 2^{-64},\label{equation:PRNG}
\end{equation}
where $x(i)=\sum_{j=-64}^{64}x(i)_j\cdot 2^j$, $x(i)_j\in\{0,1\}$,
$H(x(i))=\sum_{j=-64}^{64}\left(\bigoplus_{k=-64}^{-1}x(i)_k\right)\cdot
2^j$ and $\oplus$ denotes bitwise XOR. Then, the controlling PRBG
$(b(i))_{i=0}^{129N/15-1}$ is derived from $(x(i))_{i=0}^{N/15-1}$
by extracting the 129 bits from each $x(i)$. The above PRBG is a
special case of the second class of chaos-based PRBG proposed in
\cite{KJ:CPRBS:IEEETCASI2003}, with the parameters $p=419$, $m=8$,
$M=k=64$.

\item
\textit{The initialization process}

1) run the above PRBG to generate the controlling PRBS
$(b(i))_{i=0}^{129N/15-1}$; 2) set $temp=Secret$.

\item
\textit{The encryption procedure}

For each plain-block $f^{(15)}(k)$, do the following operations
consecutively:

\begin{itemize}
\item
\textit{Step a) Data expansion}

Add $temp$ to the 15-byte plain-block to get an expanded 16-byte
block
\[
f^{(16)}(k)=(f^{(16)}(k,j))_{j=0}^{15}=(f^{(15)}(k,0), \ldots,
f^{(15)}(k,14), temp),
\]
and then set $temp=f^{(16)}(k, l(k))$, where
$l(k)=\sum_{i=0}^3b(129k+i)\cdot 2^i$.

\item
\textit{Step b) Byte swapping}

Define a pseudorandom byte swapping operation, $Swap_{b(129k+l)}$
$(f^{(16)}(k,i),f^{(16)}(k,j))$, which swaps $f^{(16)}(k,i)$ and
$f^{(16)}(k,j)$ when $b(129k+l)=1$. Then, perform the byte swapping
operation for the following 32 values of $(i,j,l)$ one after
another: (0,8,4), (1,9,5), (2,10,6), (3,11,7), (4,12,8), (5,13,9),
(6,14,10), (7,15,11), (0,4,12), (1,5,13), (2,6,14), (3,7,15),
(8,12,16), (9,13,17), (10,14,18), (11,15,19), (0,2,20), (1,3,21),
(4,6,22), (5,7,23), (8,10,24), (9,11,25), (12,14,26), (13,15,27),
(0,1,28), (2,3,29), (4,5,30), (6,7,31), (8,9,32), (10,11,33),
(12,13,34), (14,15,35). Denote the permuted 16-byte block by
$f^{*(16)}(k)$.

\item
\textit{Step c) Value masking}

Determine two pseudo-random variables, $Seed1(k)=$
$\sum\nolimits_{i=0}^{15}\left(\bigoplus
\nolimits_{t=0}^3b(129k+4i+t)\right)\cdot 2^{i}$ and
$Seed2(k)=\sum\nolimits_{i=16}^{31}\left(\bigoplus\nolimits_{t=0}^3b(129k+4i+t)\right)\cdot
2^{i-16}$, and then do the following masking operation for $j=0 \sim
7$:
\begin{equation}
f^{**(16)}(k)_j=f^{*(16)}(k)_j\oplus
Seed(k,j),\label{equation:masking}
\end{equation}
where $f^{*(16)}(k)_j$ and $f^{**(16)}(k)_j$ are composed of the
$j$-th bits of the 16 elements of $f^{*(16)}(k)$ and
$f^{**(16)}(k)$, respectively,
\begin{equation}\label{equation:randomseed}
Seed(k,j)=
\begin{cases}
Seed1(k), & B(k,j)=3,\\
\overline{Seed1(k)}, & B(k,j)=2,\\
Seed2(k), & B(k,j)=1,\\
\overline{Seed2(k)},& B(k,j)=0,
\end{cases}
\end{equation}
and $B(k,j)=2\cdot b(129k+36+2j)+b(129k+37+2j)$.

\item
\textit{Step d) Horizontal bit rotation}

Construct an $8\times 8$ matrix $\mymatrix{M}_1$ by assigning
$\mymatrix{M}_1(i,j)$ as the $j$-th bit of $f^{**(16)}(k,i)$. Then,
perform the following horizontal bit rotation operations for
$i=0,\ldots,7$ to get a new matrix $\widetilde{\mymatrix{M}}_1$:
\[
\widetilde{\mymatrix{M}}_1(i,:)=RotateX^{p_{1,k,i},r_{1,k,i}}\left(\mymatrix{M}_1(i,:)\right),
\]
which shifts $\mymatrix{M}_1(i,:)$ (the $i$-th row of
$\mymatrix{M}_1$) by $r_{1,k,i}$ elements (bits) to the left when
$p_{1,k,i}=1$ and to the right when $p_{1,k,i}=0$. The values of the
two parameters are as follows: $p_{1,k,i}=b(129k+65+2i)$,
$r_{1,k,i}=\alpha_1+\beta_1\cdot b(129k+66+2i)$. Equivalently, the
above process can be rewritten in the following way:
\[
\widetilde{\mymatrix{M}}_1(i,:)=RotateX^{0,\overline{r}_{1,k,i}}(\mymatrix{M}_1(i,:)),
\]
where
\[
\overline{r}_{1,k,i}=\begin{cases}%
\alpha_1+\beta_1\cdot b(129k+66+2i), &
p_{1,k,i}=b(129k+65+2i)=0,\\
8-(\alpha_1+\beta_1\cdot b(129k+66+2i)), &
p_{1,k,i}=b(129k+65+2i)=1.
\end{cases}
\]
In the following, we will use the latter form to simplify our
further discussion.

In a similar way, construct another $8\times 8$ matrix
$\mymatrix{M}_2$ by assigning $\mymatrix{M}_2(i,j)$ as the $j$-th
bit of $f^{**(16)}(k,8+i)$. Then, perform similar horizontal bit
rotation operations on $\mymatrix{M}_2$ to get a new matrix
$\widetilde{\mymatrix{M}}_2$:
\[
\widetilde{\mymatrix{M}}_2(i,:)=RotateX^{0,\overline{r}_{2,k,i}}(\mymatrix{M}_2(i,:)),
\]
where
\[
\overline{r}_{2,k,i}=\begin{cases}%
\alpha_1+\beta_1\cdot b(129k+98+2i), &
p_{2,k,i}=b(129k+97+2i)=0,\\
8-(\alpha_1+\beta_1\cdot b(129k+98+2i)), &
p_{2,k,i}=b(129k+97+2i)=1.
\end{cases}
\]
After the above horizontal bit rotation operations, represent the
$i$-th byte in the 16-byte block as follows
\[
f^{\star(16)}(k,i)=\begin{cases}
\sum_{j=0}^7\widetilde{\mymatrix{M}}_1(i,j)\cdot 2^j, & 0\leq i\leq 7,\\
\sum_{j=0}^7\widetilde{\mymatrix{M}}_2(i-8,j)\cdot 2^j, & 8\leq
i\leq 15.
\end{cases}
\]

\item
\textit{Step e) Vertical bit rotation}

For $j=0,\ldots,7$, do the following vertical bit rotation
operations on $\widetilde{\mymatrix{M}}_1$ to get
$\widehat{\mymatrix{M}}_1$
\[
\widehat{\mymatrix{M}}_1(:,j)=RotateY^{0,\overline{s}_{1,k,j}}(\widetilde{\mymatrix{M}}_1(:,j)),
\]
which shifts $\widetilde{\mymatrix{M}}_1(:,j)$ (the $j$-th column of
$\widetilde{\mymatrix{M}}_1$) by $s_{1,k,j}$ elements (bits)
downwards. The value of the parameter is as follows:
\[
\overline{s}_{1,k,j}=\begin{cases}%
\alpha_1+\beta_1\cdot b(129k+82+2j), &
q_{1,k,j}=b(129k+81+2j)=0,\\
8-(\alpha_1+\beta_1\cdot b(129k+82+2j)), &
q_{1,k,j}=b(129k+81+2j)=1.
\end{cases}
\]

Similar vertical bit rotations are performed on
$\widetilde{\mymatrix{M}}_2$ to get $\widehat{\mymatrix{M}}_2$ as
follows:
\[
\widehat{\mymatrix{M}}_2(:,j)=RotateY^{0,\overline{s}_{2,k,j}}(\widetilde{\mymatrix{M}}_2(:,j)),
\]
where
\[
\overline{s}_{2,k,j}=\begin{cases}%
\alpha_1+\beta_1\cdot b(129k+114+2j), &
q_{2,k,j}=b(129k+113+2j)=0,\\
8-(\alpha_1+\beta_1\cdot b(129k+114+2j)), &
q_{2,k,j}=b(129k+113+2j)=1.
\end{cases}
\]
\color{black}

Finally, the cipher-block $f'^{(16)}(k)=(f'^{(16)}(k,i))_{i=0}^{15}$
is derived from $\widehat{\mymatrix{M}}_1$ and
$\widehat{\mymatrix{M}}_2$ as follows:
\[
f'^{(16)}(k,i)=\begin{cases}
\sum_{j=0}^7\widehat{\mymatrix{M}}_1(i,j)\cdot 2^j, & 0\leq i\leq 7,\\
\sum_{j=0}^7\widehat{\mymatrix{M}}_2(i-8,j)\cdot 2^j, & 8\leq i\leq
15.
\end{cases}
\]
\end{itemize}

\item
\textit{The decryption procedure} is simply the inverse of the above
encryption procedure.

\end{itemize}

\section{Cryptanalysis}
\label{sec:Cryptanalysis}

First of all, we point out that the subkey $Secret$ has no
influence on the decryption process. It is because $Secret$ is only
used to determine the expanded byte, and never used to change the
value of any other byte in the plaintext. In fact, if we use a
different value of $Secret$ for the decryption process, the
plaintext can still be correctly recovered. Furthermore, the
probability that $Secret$ becomes the expanded byte of $f^{(16)}(k)$
is $(15/16)^k$, which decreases very exponentially. This means that
$Secret$ has no influence on the encryption process after $k$ become
sufficiently large. As a whole, $Secret$ should be excluded from the
key. In the rest of this paper, we will not consider $Secret$ as a
subkey.

\subsection{Some properties of MCS}

Define the XOR-differential (``differential" in short hereinafter)
of two plaintexts $f_0$ and $f_1$ as $f_{0\oplus 1}=f_0\oplus f_1$.
When $f_0$ and $f_1$ are encrypted with the same secret key, it is
easy to prove the following three properties of MES, which will be
the basis of the proposed attack.

\begin{property}
The random masking in Step \textit{c)} cannot change the
differential value, i.e., $\forall\; k,j$,
$f^{**(16)}_{0\oplus1}(k,j)\equiv f^{*(16)}_{0\oplus1}(k,j)$.
\end{property}
\begin{proof}
It is a straightforward result of the following property of XOR:
$(a\oplus x)\oplus(b\oplus x)=a\oplus b$.
\end{proof}

\begin{property}
Each expanded plain-block $f_{0\oplus 1}^{(16)}(k)$ is independent
of the sub-key $Secret$.
\end{property}
\begin{proof}
This can be proved with mathematical induction on $k$. When $k=0$
and $0\leq j\leq 15$, i.e., for the $j$-th byte of the first 16-byte
block,
\[
f_{0\oplus 1}^{(16)}(0,j)=\begin{cases}%
f_{0\oplus 1}^{(15)}(0,j), & 0\leq j\leq 14,\\
Secret\oplus Secret=0, & j=15,
\end{cases}
\]
which is obviously independent of the value of $Secret$. Now assume
the property holds for the first $k-1$ blocks. Then, for the $k$-th
16-byte block,
\[
f_{0\oplus 1}^{(16)}(k,j)=\begin{cases}%
f_{0\oplus 1}^{(15)}(k,j), & 0\leq j\leq 14,\\
f_{0\oplus 1}^{(16)}(k-1,l(k-1)), & j=15,\\
\end{cases}
\]
which is also independent from $Secret$ according to the assumption.
Thus, this property is proved.
\end{proof}

\begin{property}
The byte swapping in \textit{Step b)} cannot change each
differential value, but its position in the 16-byte block.
\end{property}

\begin{property}
Both the horizontal bit rotation in \textit{Step d)} and the
vertical bit rotation in \textit{Step e)} cannot change each
differential bit itself, but its position in the binary presentation
of the 8-byte block.
\end{property}

The proofs of the above two properties are straightforward, so we
omit them here.

\subsection{The differential attack}

Based on the above properties of MCS, the data expansion in
\textit{Step a)}, the first eight byte swapping operations in
\textit{Step b)}, the vertical bit rotation in \textit{Step e)}, the
horizontal bit rotation in \textit{Step d)}, the other unkown byte
swapping operations in \textit{Step b)} and the value masking in
\textit{Step c)} can be broken in order with a number of chosen
plaintext differentials.

\subsubsection{Breaking the secret data expansion in Step a)}
\label{sssec:breakingDataExpan}

\newcommand\HammingWeight[1]{\left|#1\right|}

To facilitate the following discussion, let us denote the Hamming
weight of a byte or a block $x$, i.e., the number of 1-bits in $x$,
by $\HammingWeight{x}$. From Property 2, one can see that there are
$8\cdot 15 =120$ binary bits of $f_{0\oplus 1}'^{(16)}(k)$ come from
$f_{0\oplus 1}^{(15)}(k)$ and other eight bits come from $f_{0\oplus
1}^{(15)}(k-1, l(k-1))$ for $k\geq 1$ (the eight expanded bits are
all 0-bits when $k=0$). Since all the other steps do not change the
Hamming weight of each 16-byte block, we can get
$\HammingWeight{f_{0\oplus 1}^{(15)}(k-1,
l(k-1))}=\HammingWeight{f_{0\oplus
1}^{'(16)}(k)}-\HammingWeight{f_{0\oplus 1}^{(15)}(k)}$. In case
$\HammingWeight{f_{0\oplus 1}^{(15)}(k-1, l(k-1))}$ is unique in the
last 15-byte block $f_{0\oplus 1}^{(15)}(k-1)$, we can uniquely
determine the value of $l(k-1)$. Considering
$\HammingWeight{f_{0\oplus 1}^{(15)}(k-1, l(k-1))}\in\{0,\ldots,8\}$
but $l(k-1)\in\{0,\ldots,15\}$, at least two plain-bytes in each
15-byte block have the same Hamming weight. So, the value of
$l(k-1)$ may not be uniquely determined sometimes. To make the
unique determination of $l(k-1)$ possible, we can choose two
plaintext differentials $f_{0\oplus 1}$ and $f_{0\oplus 2}$ (i.e.,
differentials of three chosen plaintexts $f_0$, $f_1$ and $f_2$) to
fulfill the following two requirements: 1) $\forall k,j_1\neq j_2$,
$\left(\HammingWeight{f_{0\oplus
1}^{(15)}(k,j_1)},\HammingWeight{f_{0\oplus
2}^{(15)}(k,j_1)}\right)\neq\left(\HammingWeight{f_{0\oplus
1}^{(15)}(k,j_2)},\HammingWeight{f_{0\oplus
2}^{(15)}(k,j_2)}\right)$; 2) $\forall k, j$,
$\left(\HammingWeight{f_{0\oplus
1}^{(15)}(k,j)},\HammingWeight{f_{0\oplus
2}^{(15)}(k,j)}\right)\neq(0,0)$. For example, the two plaintext
differentials can be chosen to have the following Hamming weights:
\begin{eqnarray*}
\left(\HammingWeight{f_{0\oplus 1}(i)}\right)_{i=1}^{N-1} & = &
(\overbrace{0,0,0,0,0,0,0,0,1,1,1,1,1,1,1,1,1,\ldots,8,8,8,8,8,8,8,8,8}^{9\times9-1=80\text{ elements}},\ldots)\\
\left(\HammingWeight{f_{0\oplus 2}(i)}\right)_{i=1}^{N-1} & = &
(1,2,3,4,5,6,7,8,0,1,2,3,4,5,6,7,8,\ldots,0,1,2,3,4,5,6,7,8,\ldots)
\end{eqnarray*}
With the above chosen plaintexts, it is obvious that the value of
$l(k-1)$ can always be uniquely determined, except when
\begin{equation}
\left(\HammingWeight{f_{0\oplus
1}^{(15)}(k-1,15)},\HammingWeight{f_{0\oplus
2}^{(15)}(k-1,15)}\right)\in\bigcup_{j=0}^{14}\left(\HammingWeight{f_{0\oplus
1}^{(15)}(k-1,j)},\HammingWeight{f_{0\oplus
2}^{(15)}(k-1,j)}\right).
\end{equation}
We can calculate the occurrence probability of the above equation is
less than
$\frac{15}{16}\cdot\left(\frac{1}{16}\right)^{\lfloor80/15\rfloor-1}\approx
1.4305\times 10^{-5}$. For a $512\times 512$ image, this means that
we will not be able to uniquely determine the value of $l(k-1)$ for
less than $1.4305\times 10^{-5}\times 512\times 512/16\approx
0.2344$ blocks in an average sense. In other words, the value of
$l(k-1)$ can be uniquely determined for almost all blocks.
Note that breaking $l(k-1)$ implies breaking 4 controlling
bits $(b(129(k-1)+i)))_{i=0}^3$.

\subsubsection{Breaking the first eight byte-swapping operations in Step b)}
\label{sssec:breakingbits}

From Properties 3, 4, one can see that all the $8\cdot16=128$ bits
of each 16-byte expanded plain-block $f_{0\oplus 1}^{(16)}(k)$ are
the same as the ones of the corresponding 16-byte cipher-block
$f_{0\oplus 1}'^{(16)}(k)$, except that their locations may change.
Observing how the bit locations are changed in the whole encryption
process, we can see the following eight byte-swapping operations are
the only encryption operations moving bits from one 8-byte
half-block to another:
$Swap_{b(129k+i+4)}(f^{(16)}(k,i),f^{(16)}(k,i+8))$, when
$i=0,1,2,3,4,5,6,7$. Apparently, when the controlling bit is 1, each
byte-swapping operation swaps the locations of one byte in the first
half-block and the other byte in another half-block. This fact means
that, by choosing the differences between the Hamming weights of the
eight bytes in the two half-blocks properly, we will be able to
derive the values of the controlling bits $(b(129k+i+4))_{i=0}^7$.
The simplest tactic is to choose $f_{0\oplus 1}^{(16)}(k)$ such that
each half-block has only one byte with a different Hamming weight
from the corresponding byte in the other half-block. If we assume
all the values of $(l(k))_{k=0}^{N/15-2}$ have been recovered, which
happens with high probability as we shown in the previous
subsection, the first 15 bytes in $f_{0\oplus 1}^{(16)}(k)$ can be
freely chosen by choosing $f_{0\oplus 1}^{(15)}(k)$. The last byte
in each 16-byte block $f_{0\oplus 1}^{(16)}(k,15)$ may not be
chosen, if it is equal to $Secret$. Fortunately, this has no
influence on the process of breaking the first eight byte-swapping
operations, because what is chosen for the last byte is
$\HammingWeight{f^{(16)}(k,15)}-\HammingWeight{f^{(16)}(k,7)}$.
Although we may not be able to choose $f_{0\oplus 1}^{(16)}(k,15)$,
we can always choose $f_{0\oplus 1}^{(16)}(k,7)$ to have a different
Hamming weight from that of $f_{0\oplus 1}^{(16)}(k,7)$. One
chosen-block $f_{0\oplus 1}^{(16)}(k)$ will be able to derive the
value of one controlling bit, which controls the possible swapping
of the two bytes (in two half-blocks, respectively) with different
Hamming weights. We need eight chosen plain-blocks (thus eight
chosen plaintext differentials) to determine the values of all the
eight controlling bits.

While eight chosen plaintext differentials are enough to recover all
the bits controlling the first eight byte-swapping operations, we
actually need only two chosen plaintext differentials to achieve
this goal. To see how it is possible, denote the difference between
the Hamming weights of the two half-blocks of the $k$-th
cipher-block by $\Delta\HammingWeight{\left(f_{0\oplus
1}'^{(16)}(k)\right)_{i=0}^7}$. Then, we have the following
equation:
\begin{eqnarray*}
\Delta\HammingWeight{\left(f_{0\oplus 1}'^{(16)}(k)\right)_{i=0}^7}
& = & \HammingWeight{\left(f_{0\oplus
1}'^{(16)}(k,i)\right)_{i=0}^7}-\HammingWeight{\left(f_{0\oplus
1}'^{(16)}(k,i+8)\right)_{i=0}^7}\\
& = & \HammingWeight{\left(f_{0\oplus
1}^{*(16)}(k,i)\right)_{i=0}^7}-\HammingWeight{\left(f_{0\oplus
1}^{*(16)}(k,i+8)\right)_{i=0}^7}\\
& = & \sum_{i=0}^7\left(\HammingWeight{f_{0\oplus
1}^{*(16)}(k,i)}-\HammingWeight{f_{0\oplus
1}^{*(16)}(k,i+8)}\right)\\
& = & \sum_{i=0}^7b^{\pm}(k,i)\left(\HammingWeight{f_{0\oplus
1}^{(16)}(k,i)}-\HammingWeight{f_{0\oplus 1}^{(16)}(k,i+8)}\right),
\end{eqnarray*}
where
\[
b^{\pm}(k,i)=1-2b(129k+i+4)=\begin{cases}
1, & b(129k+i+4)=0,\\
-1, & b(129k+i+4)=1.
\end{cases}
\]
By choosing the values of $\left(\HammingWeight{f_{0\oplus
1}^{(16)}(k,i)}-\HammingWeight{f_{0\oplus
1}^{(16)}(k,i+8)}\right)_{i=0}^7$ to be a set of numbers such that
every nonzero number can not be represented as a linear combination
of other numbers in the set, the controlling bits corresponding to
the nonzero numbers can be determined uniquely. For instance, to
determine the values of $b^{\pm}(k,0),\ldots,b^{\pm}(k,3)$, we can
choose a plaintext differential such that
\begin{itemize}
\item
$\HammingWeight{f_{0\oplus 1}^{(16)}(k,i)}-\HammingWeight{f_{0\oplus
1}^{(16)}(k,i+8)}=\pm4,\pm5,\pm6,\pm8$ for $i=0,1,2,3$,
respectively;

\item
$\HammingWeight{f_{0\oplus 1}^{(16)}(k,i)}-\HammingWeight{f_{0\oplus
1}^{(16)}(k,i+8)}=0$ for $i=4,5,6,7$.
\end{itemize}
The above chosen plaintext differential leads to the following
result:
\[
\Delta\HammingWeight{\left(f_{0\oplus
1}^{'(16)}(k)\right)_{i=0}^7}\in\{\pm23,\pm15,\pm13,\pm11,\pm7,\pm5,\pm
3,\pm1\}.
\]
The 16 possible values of $\Delta\HammingWeight{\left(f_{0\oplus
1}^{'(16)}(k)\right)_{i=0}^7}$ correspond to the 16 possible values
of $(b(129k+4+i))_{i=0}^3$. Choosing another plaintext differential
such that
\begin{itemize}
\item
$\HammingWeight{f_{0\oplus 1}^{(16)}(k,i)}-\HammingWeight{f_{0\oplus
1}^{(16)}(k,i+8)}=0$ for $i=0,1,2,3$;

\item
$\HammingWeight{f_{0\oplus 1}^{(16)}(k,i)}-\HammingWeight{f_{0\oplus
1}^{(16)}(k,i+8)}=\pm4,\pm5,\pm6,\pm8$ for $i=4,5,6,7$,
respectively,
\end{itemize}
we will be able to uniquely determine the other four controlling
bits $(b(129k+4+i))_{i=4}^7$. As a whole, with only two chosen
plaintext differentials, we can uniquely determine all the eight
controlling bits $(b(129k+4+i))_{i=0}^7$.

\subsubsection{Breaking the other part of MCS}
\label{sssec:breakingbyteswap}

For the $k$-th block, denote the intermediate result of the first
eight byte-swapping operations by $\overline{f_{0\oplus
1}^{*(16)}}(k)$. Knowing $b(129k+4)\sim b(129k+11)$ allows us to
choose $\overline{f_{0\oplus 1}^{*(16)}}(k)$ by manipulating
$f_{0\oplus 1}^{(16)}(k)$. The other encryption operations to be
further broken include the 9th to 35th byte-swapping operations, the
value masking, and the horizontal/vertical bit rotations.

Different from the first 8 byte-swapping operations, the 9th to 35th
ones in \textit{Step b)} only shuffle the locations of the eight
bytes inside each half-block. We found these byte-swapping
operations cannot be uniquely determined, because some equivalent
but different encryption operations exist. Roughly speaking, if we
add an overall circularly byte shift operation to \textit{Step b)}
and all the other steps afterwards, we will get an encryption scheme
equivalent to but different from the real one. Therefore, in this
sub-subsection we turn to find such an equivalent encryption scheme.
To facilitate our discussion, in the following, we use the acronym
``EES'' to denote the equivalent encryption scheme that has the same
encryption performance as all the four kinds of encryption
operations to be further broken. The EES is also composed of four
parts, which correspond to the four different kinds of encryption
operations, respectively. Once again, we use a divide-and-conquer
tactic to get all the four pars of an EES.

\paragraph{Obtaining the vertical bit-rotation
part of the EES} \label{section:EBRS:VBR}

To get the vertical bit-rotation part, we need to cancel the
horizontal bit-rotation part and the byte-swapping part. The
horizontal bit rotations can be done by choosing all bytes in
$\overline{f_{0\oplus 1}^{*(16)}}(k)$ to be either 0 or 255, i.e.,
all the bits in $M_1$ and $M_2$ are identical (either 0 or 1). The
byte-swapping operations cannot be fully canceled. To minimize its
interference with the vertical bit-rotation part, we can choose each
half-block such that there is only one 0 or one 255. Without loss of
generality, we choose one plaintext differential such that both
half-blocks of each 16-byte block $\overline{f_{0\oplus
1}^{*(16)}}(k)$ contains only one 255-byte but seven 0-bytes, i.e.,
\[
\left(\overline{f_{0\oplus
1}^{*(16)}}(k,i)\right)_{i=0}^7=\left(\overline{f_{0\oplus
1}^{*(16)}}(k,i)\right)_{i=8}^{15}=(\overbrace{0,\ldots,0}^{l\text{
zeros}},255,0,\ldots,0).
\]
After the byte-swapping operations, assume $\overline{f_{0\oplus
1}^{*(16)}}(k,l)$ is moved to $f_{0\oplus
1}^{*(16)}(k,\widetilde{s}_{1,k,l})$ and $\overline{f_{0\oplus
1}^{*(16)}}(k,8+l)$ to $f_{0\oplus
1}^{*(16)}(k,8+\widetilde{s}_{2,k,l})$, where
$\widetilde{s}_{1,k,l},\widetilde{s}_{2,k,l}\in\{0,\ldots,7\}$.
Since the horizontal bit rotations are canceled, by comparing
$(\overline{f_{0\oplus 1}^{*(16)}}(k,i))_{i=0}^7$ and
$(\overline{f_{0\oplus 1}'^{(16)}}(k,i))_{i=0}^7$, we can observe
that $RotateY^{0,\overline{s}_{1,k,j}+\widetilde{s}_{1,k,l}}$ is
performed for the $j$-th bit of $\overline{f_{0\oplus
1}^{*(16)}}(k,0)$. Similarly, for the second half-block, we can
observe that
$RotateY^{0,\overline{s}_{2,k,j}+\widetilde{s}_{2,k,l}}$ is
performed for the $j$-th bit of $\overline{f_{0\oplus
1}^{*(16)}}(k,8)$.

\paragraph{Obtaining the horizontal bit-rotation part of the EES}

Now, we need to cancel the byte-swapping operations and the vertical
bit rotations. The byte-swapping operations can be canceled by
choosing a second plaintext differential such that all the bytes in
each half-block are identical. To distinguish the horizontal bit
shifts, we should choose the byte $x\in\{0,\ldots, 255\}$ to satisfy
the following property: $a_1\not\equiv a_2\pmod
8\Leftrightarrow(x\ggg a_1)\neq (x\ggg a_2)$, or equivalently,
$a_1\equiv a_2\pmod 8\Leftrightarrow(x\ggg a_1)=(x\ggg a_2)$. The
simplest choice of $x$ is $2^i$, where $i\in\{0,\ldots,7\}$. When
$f^{(16)}(k,15)=temp$, either $\overline{f_{0\oplus
1}^{*(16)}}(k,7)$ or $\overline{f_{0\oplus 1}^{*(16)}}(k,15)$ will
always be 0, so it will not be possible to obtain the horizontal
bit-rotation part for this byte. Fortunately, this does not
influence the decryption process, because the expanded byte is
actually redundant and will be finally discarded. The vertical bit
rotations cannot be canceled, since they are performed after the
horizontal bit rotations. Since we have obtained the vertical
bit-rotation part of the EES, we can apply it to $(f_{0\oplus
1}'^{(16)}(k,i))_{i=0}^7$ to get $(f_{0\oplus
1}^{\star(16)}(k,i\dotplus \widetilde{s}_{1,k,l}))_{i=0}^7$, where
$\dotplus$ denotes addition modulus 8. Then, compare $(f_{0\oplus
1}^{\star(16)}(k,i\dotplus \widetilde{s}_{1,k,l}))_{i=0}^7$ with
$(\overline{f_{0\oplus 1}^{*(16)}}(k,i))_{i=0}^7$, one can observe
that $RotateX^{0,\bar{r}_{1,k,(i\dotplus \widetilde{s}_{1,k,l})}}$
is performed for $\overline{f_{0\oplus 1}^{*(16)}}(k,i)$. Similarly,
we can observe $RotateX^{0,\bar{r}_{2,k,(i\dotplus
\widetilde{s}_{2,k,l})}}$ is performed for $\overline{f_{0\oplus
1}^{*(16)}}(k,8+i)$.

\paragraph{Obtaining the byte-swapping part of the EES}
\label{section:EBRS:BS}

After obtaining the horizontal/vertical bit-rotation parts of the
EES, we can apply the inverse horizontal/vertical bit rotations to
$(f_{0\oplus 1}'^{(16)}(k, j))_{j=0}^{15}$ to get $(f_{0\oplus
1}^{*(16)}(k,\widetilde{s}_{1,k,l}\dotplus i))_{i=0}^{7}$ and
$(f_{0\oplus 1}^{*(16)}(k,8+(\widetilde{s}_{2,k,l}\dotplus
i)))_{i=0}^{7}$. If we choose $f_{0\oplus 1}^{*(16)}(k)$ such that
all the eight bytes of each half-block are different from each
other, we will be able to obtain the following byte-swapping part of
the EES. For the first half-block, the real byte-swapping operation
moves $\overline{f_{0\oplus 1}^{*(16)}}(k,i)$ to $f_{0\oplus
1}^{*(16)}(k,\widetilde{s}_{1,k,i})$, the one we obtained for the
EES will move it to $f_{0\oplus
1}^{*(16)}(k,\widetilde{s}_{1,k,i}\dot{-}\widetilde{s}_{1,k,l})$,
where $\dot{-}$ denotes subtraction modulus 8. Similarly, for the
second half-block, the real byte-swapping operation moves
$\overline{f_{0\oplus 1}^{*(16)}}(k,8+i)$ to $f_{0\oplus
1}^{*(16)}(k,8+\widetilde{s}_{2,k,i})$, the one we obtained for the
EES will move it to $f_{0\oplus
1}^{*(16)}(k,8+(\widetilde{s}_{2,k,i}\dot{-}\widetilde{s}_{2,k,l}))$.

\paragraph{Obtaining the value-masking part of the EES}

After obtaining the byte-swapping part of the EES, we can get
$\{f^{*(16)}(k,i\dotplus\widetilde{s}_{1,k,l})\}_{i=0}^7$ and
$\{f^{*(16)}(k,8+(i\dotplus\widetilde{s}_{1,k,l}))\}_{i=0}^7$ from
any known plaintext. In addition, after obtaining both the
horizontal and vertical bit-rotation parts, we can get
$\{f^{**(16)}(k,i\dotplus\widetilde{s}_{1,k,l})\}_{i=0}^7$ and
$\{f^{**(16)}(k,8+(i\dotplus\widetilde{s}_{1,k,l}))\}_{i=0}^7$ from
any known ciphertext. We do not need to choose more plaintexts, but
can simply reuse any chosen plaintext used in previous steps. Note
that the value masking performed in \textit{Step c)} can be
rewritten as the equivalent form: for $i=0,\ldots,15$,
\begin{equation}
f^{**(16)}(k,i)=f^{*(16)}(k,i)\oplus
Seed^*(k,i),\label{equation:masking2}
\end{equation}
where $Seed^*(k,i)=\sum_{j=0}^7 Seed(k,j)_i\cdot 2^j$ and
$Seed(k,j)_i$ is the $i$-th bit of $Seed(k,j)$. Then, by XORing
$\{f^{*(16)}(k,i\dotplus\widetilde{s}_{1,k,l})\}_{i=0}^7$ and
$\{f^{**(16)}(k,i\dotplus\widetilde{s}_{1,k,l})\}_{i=0}^7$, we can
get $(Seed^*(k,i\dotplus\widetilde{s}_{1,k,l}))_{i=0}^7$. Similarly,
by XORing
$\{f^{*(16)}(k,8+(i\dotplus\widetilde{s}_{1,k,l}))\}_{i=0}^7$ and
$\{f^{**(16)}(k,8+(i\dotplus\widetilde{s}_{1,k,l}))\}_{i=0}^7$, we
can get $(Seed^*(k,8+(i\dotplus\widetilde{s}_{1,k,l})))_{i=0}^7$.

Observing the above four results, we can see all the fours parts of
the ESS are related to the unknown parameters
$\widetilde{s}_{1,k,l}$ and $\widetilde{s}_{2,k,l}$. If we choose
different value of $l$ in Sec.~\ref{section:EBRS:VBR}, we may have
different ESS. All the possible EESs are equivalent to each other
(and to the real encryption scheme), so we can use any of them to
decrypt any ciphertext encrypted with the same key, as long as the
size of the ciphertext is not larger than $N$. In the next
subsection, we will show the values of $\widetilde{s}_{1,k,l}$ and
$\widetilde{s}_{2,k,l}$ can be uniquely determined if the sub-keys
$\alpha_1$, $\alpha_2$, $\beta_1$ and $\beta_2$ satisfy some
requirements.

\subsubsection{Performance of the differential attack}
\label{section:AttackPerformance}

To sum up, the differential attack outputs the following items as an
equivalent key:
\begin{itemize}
\item
for data expansion: $(l(k-1))_{1\leq k\leq N/15-1}$, which is
equivalent to $(b(129(k-1)+i))_{1\leq k\leq N/15-1 \atop 0\leq i\leq
3}$;

\item
for the first eight byte-swapping operations: $(b(129k+i))_{0\leq
k\leq N/15-1 \atop 4\leq i\leq 11}$;

\item
for the 9th to 35th byte-swapping operations:
$\left\{\overline{f_{0\oplus 1}^{*(16)}}(k,i)\to f_{0\oplus
1}^{*(16)}(k,\widetilde{s}_{1,k,i}\dot{-}\widetilde{s}_{1,k,l})\right\}_{0\leq
k\leq N/15-1 \atop 0\leq i\leq 7}$ and $\left\{\overline{f_{0\oplus
1}^{*(16)}}(k,8+i)\to f_{0\oplus
1}^{*(16)}(k,8+(\widetilde{s}_{2,k,i}\dot{-}\widetilde{s}_{2,k,l}))\right\}_{0\leq
k\leq N/15-1 \atop 0\leq i\leq 7}$;

\item
for the value masking:
$\left(Seed^*(k,(i\dotplus\widetilde{s}_{1,k,l}))\right)_{0\leq
k\leq N/15-1 \atop 0\leq i\leq 7}$ and
$\left(Seed^*(k,8+(i\dotplus\widetilde{s}_{1,k,l}))\right)_{0\leq
k\leq N/15-1 \atop 0\leq i\leq 7}$;

\item
for the horizontal bit rotations:
$\left(RotateX^{0,\overline{r}_{1,k,(i\dotplus
\widetilde{s}_{1,k,l})}}\right)_{0\leq k\leq N/15-1 \atop 0\leq
j\leq 7}$ and $\left(RotateX^{0,\bar{r}_{2,k,(i\dotplus
\widetilde{s}_{2,k,l})}}\right)_{0\leq k\leq N/15-1 \atop 0\leq
j\leq 7}$;

\item
for the vertical bit rotations:
$\left(RotateY^{0,\overline{s}_{1,k,j}+\widetilde{s}_{1,k,l}}\right)_{0\leq
k\leq N/15-1 \atop 0\leq j\leq 7}$ and
$\left(RotateY^{0,\overline{s}_{2,k,j}+\widetilde{s}_{2,k,l}}\right)_{0\leq
k\leq N/15-1 \atop 0\leq j\leq 7}$.
\end{itemize}
All the above items form an encryption system equivalent to MCS and
can be used to decrypt any ciphertexts encrypted with the same
secret key. The (equivalent) encryption operations performed on some
expanded bytes $f^{(16)}(k,15)$ may not be recovered, but which does
not influence the effectiveness of the differential attack, since
those expanded bytes will finally be discarded.

The total number of chosen plaintexts is the sum of the following:
a) two differentials for breaking the data expansion; b) two
differentials for breaking the first eight byte-swapping operations;
c) four differentials for obtaining the EES. Note that the plaintext
differential needed in Sec.~\ref{section:EBRS:BS} can be replaced by
the two differentials in Sec.~\ref{sssec:breakingDataExpan}. So, we
only need two more differentials for obtaining the EES. As a whole,
the differential attack requires $2+2+2=6$ plaintext differentials,
or seven plaintexts, to break MCS.

The complexity of the differential attack is also very small, since
we do not have any exhaustive search process in all the steps
described above. With 6 chosen plaintext differentials, the
computational complexity of the attack is just $O(6N)=O(N)$, which
is the same as that of the normal encryption/decryption process of
MCS.

\subsection{Breaking some sub-keys and more controlling bits}

The differential attack described in the previous subsection outputs
an equivalent key, which include some controlling bits
$(b(129k+i))_{i=0}^{11}$, but does not include any part of the
secret key. In this subsection, we show we may further derive more
controlling bits and the following four sub-keys: $\alpha_1$,
$\beta_1$, $\alpha_2$ and $\beta_2$. Although we have not found a
way to break the underlying pseudorandom bit generator (PRBG) and
then break the subkey $x(0)$, breaking more controlling bits makes
it easier to analyze more potential weaknesses of the PRBG and opens
the door to a successful cryptanalysis in future.

We first try to break the two sets
$\mathbb{R}_1=\{\alpha_1,8-\alpha_1,\alpha_1+\beta_1,8-(\alpha_1+\beta_1)\}$
and
$\mathbb{R}_2=\{\alpha_2,8-\alpha_2,\alpha_2+\beta_2,8-(\alpha_2+\beta_2)\}$.
Then, we may be able to further determine sub-keys
$\alpha_1,\beta_1,\alpha_2,\beta_2$, $\widetilde{s}_{1,k,l}$,
$\widetilde{s}_{2,k,l}$, and more controlling bits.

\subsubsection{Breaking $\mathbb{R}_1$ and $\mathbb{R}_2$}

In the differential attack, what we have obtained for the horizontal
bit rotations are $\left(RotateX^{0,\overline{r}_{1,k,(i\dotplus
\widetilde{s}_{1,k,l})}}\right)_{0\leq k\leq N/15-1 \atop 0\leq
j\leq 7}$ and $\left(RotateX^{0,\bar{r}_{2,k,(i\dotplus
\widetilde{s}_{2,k,l})}}\right)_{0\leq k\leq N/15-1 \atop 0\leq
j\leq 7}$. According to how $\overline{r}_{1,k,i}$ and
$\overline{r}_{1,k,i}$ are determined, it is obvious that
$\mathbb{R}_{1,k}=\{\overline{r}_{1,k,(i\dotplus
\widetilde{s}_{1,k,l})}\}_{i=0}^7\subseteq\mathbb{R}_1$ and
$\mathbb{R}_{2,k}=\{\overline{r}_{2,k,(i\dotplus
\widetilde{s}_{2,k,l})}\}_{i=0}^7\subseteq\mathbb{R}_2$. Assuming
the secret bits controlling $(\overline{r}_{1,k,i})_{i=0}^7$ and
$(\overline{r}_{2,k,i})_{i=0}^7$ distribute uniformly over
$\{0,1\}$, from Proposition~\ref{proposition:ProbX} we can get
\[
\text{Prob}\left(\mathbb{R}_1\neq\bigcup_{0\leq k\leq N/15-1 \atop
0\leq i\leq 7}\left\{\overline{r}_{1,k,(i\dotplus
\widetilde{s}_{1,k,l})},8-\overline{r}_{1,k,(i\dotplus
\widetilde{s}_{1,k,l})}\right\}\right)\leq 2/2^{8N/15}=1/2^{8N/15-1}
\]
and
\[
\text{Prob}\left(\mathbb{R}_2\neq\bigcup_{0\leq k\leq N/15-1 \atop
0\leq i\leq 7}\left\{\overline{r}_{2,k,(i\dotplus
\widetilde{s}_{2,k,l})},8-\overline{r}_{2,k,(i\dotplus
\widetilde{s}_{2,k,l})}\right\}\right)\leq 1/2^{8N/15-1}.
\]
Since $8N/15-1$ is generally very large, the above two probability
is extremely small, which means that $\mathbb{R}_1$ and
$\mathbb{R}_2$ can be uniquely determined with very high
probability.

\begin{proposition}
\label{proposition:ProbX}Assume $1\leq\beta\leq 7$,
$1\leq\alpha<\alpha+\beta\leq 7$ and
$\mathbb{R}=\{\alpha,8-\alpha,\alpha+\beta,8-(\alpha+\beta)\}$. If
for $i=1,\ldots,n$, random variable $r_i\in\mathbb{Z}$ satisfies
$\text{Prob}(r_i\in\{\alpha,8-\alpha\})=p$, then
\[
\text{Prob}\left(\mathbb{R}\neq\bigcup_{i=1}^n\{r_i,8-r_i\}\right)=\begin{cases}%
0, & 2\alpha+\beta=8,\\
1, & 2\alpha+\beta\neq 8\text{ and }n=1,\\
p^n+(1-p)^n, & 2\alpha+\beta\neq 8\text{ and }n\geq 2.\end{cases}
\]
\end{proposition}
\begin{proof}
When $2\alpha+\beta=8$, we can get $\alpha=8-(\alpha+\beta)$ and
$8-\alpha=\alpha+\beta$, which leads to
$\mathbb{R}=\{\alpha,8-\alpha\}=\{\alpha+\beta,8-(\alpha+\beta)\}$.
Hence, we can immediately get $\{r_i,8-r_i\}=\mathbb{R}$ and then
$\bigcup_{i=1}^n\{r_i,8-r_i\}=\mathbb{R}$. This means that
$\text{Prob}\left(\mathbb{R}\neq\bigcup_{i=1}^n\{r_i,8-r_i\}\right)=0$.

When $2\alpha+\beta\neq 8$, we have $\alpha\neq 8-(\alpha+\beta)$
and $8-\alpha\neq\alpha+\beta$. Since $\alpha\neq\alpha+8$ and
$8-\alpha\neq 8-(\alpha+\beta)$, there are only the following
$\binom{4}{2}-4=2$ pairs of elements that may be equal to each other
to make $\#(\mathbb{R})<4$, where $\#(\cdot)$ denotes the
cardinality of a set:
\begin{itemize}
\item
$\alpha=8-\alpha$: $\alpha=4\Rightarrow 1\leq\beta\leq 3$ and
$\mathbb{R}=\{4,4,4+\beta,4-\beta\}\Rightarrow\#(\mathbb{R})=3$;

\item
$\alpha+\beta=8-(\alpha+\beta)$: $\alpha+\beta=4\Rightarrow
1\leq\alpha\leq 3$ and
$\mathbb{R}=\{\alpha,8-\alpha,4,4\}\Rightarrow\#(\mathbb{R})=3$.
\end{itemize}
In case no any two elements in $\mathbb{R}$ are equal to each other,
it is obvious that $\#(\mathbb{R})=4$. As a whole, we have
$\#(\mathbb{R})\geq 3$. Then, when $n=1$, the proposition is
obviously true since $\#(\{r_i,8-r_i\})<3\leq\#(\mathbb{R})$. When
$n\geq 2$, we can see there are only two ways to make
$\mathbb{R}\neq\bigcup_{i=1}^n\{r_i,8-r_i\}$:
\begin{itemize}
\item
$\bigcup_{i=1}^n\{r_i,8-r_i\}=\{\alpha,8-\alpha\}$, which occurs
with probability $p^n$;

\item
$\bigcup_{i=1}^n\{r_i,8-r_i\}=\{\alpha+\beta,8-(\alpha+\beta)\}$,
which occurs with probability $(1-p)^n$.
\end{itemize}
As a whole, we have
$\text{Prob}(\mathbb{R}\neq\bigcup_{i=1}^n\{r_i,8-r_i\})=p^n+(1-p)^n$.

Combining the above three different cases, the proposition is thus
proved.
\end{proof}

\subsubsection{Determining sub-keys $\alpha_1$, $\beta_1$, $\alpha_2$ and $\beta_2$}
\label{section:BreakingAlphaBeta12}

After getting $\mathbb{R}_1$ and $\mathbb{R}_2$, the four sub-keys
$\alpha_1$, $\beta_1$, $\alpha_2$ and $\beta_2$ may be uniquely
determined. Following a similar process of the proof of
Proposition~\ref{proposition:ProbX}, we consider the following three
cases for $m=1,2$:
\begin{itemize}
\item
$\#(\mathbb{R}_m)=2$: This case happens only when
$2\alpha_m+\beta_m=8$. There are three possible sets
$\mathbb{R}_m=\{1,7\},\{2,6\},\{3,5\}$, which corresponds to
$(\alpha_m,\beta_m)=(1,6),(2,4),(3,2)$, respectively. Apparently,
knowing $\mathbb{R}_m$ allows us to uniquely determine the values of
$\alpha_m$ and $\beta_m$.

\item
$\#(\mathbb{R}_m)=3$: This case happens when $\alpha_m=8-\alpha_m=4$
or $\alpha_m+\beta_m=8-(\alpha_m+\beta_m)=4$. There are only three
possible sets $\mathbb{R}_m$, each of which corresponds to two
possible values of $(\alpha_m,\beta_m)$:
\begin{itemize}
\item
$\mathbb{R}_m=\{4,1,7\}$: $(\alpha_m,\beta_m)=(4,3)$ or (1,3);

\item
$\mathbb{R}_m=\{4,2,6\}$: $(\alpha_m,\beta_m)=(4,2)$ or (2,2);

\item
$\mathbb{R}_m=\{4,3,5\}$: $(\alpha_m,\beta_m)=(4,1)$ or (3,1).
\end{itemize}
It can be seen that $\alpha_m$ and $\beta_m$ cannot be uniquely
determined in this case.

\item
$\#(\mathbb{R}_m)=4$: This case includes three possible sets
$\mathbb{R}_m$, each of which corresponds to four different values
of $(\alpha_m,\beta_m)$:
\begin{itemize}
\item
$\mathbb{R}_m=\{1,2,6,7\}$: $(\alpha_m,\beta_m)=(1,1)$, (1,5), (2,5)
or (6,1);

\item
$\mathbb{R}_m=\{1,3,5,7\}$: $(\alpha_m,\beta_m)=(1,2)$, (1,4), (3,4)
or (5,2);

\item
$\mathbb{R}_m=\{2,3,5,6\}$: $(\alpha_m,\beta_m)=(2,1)$, (2,3), (3,3)
or (5,1).
\end{itemize}
\end{itemize}

\subsubsection{Determining $\widetilde{s}_{1,k,l}$ and $\widetilde{s}_{2,k,l}$}
\label{section:BreakingVBR2}

In the differential attack, what we have obtained for the vertical
bit rotations are
$\left(RotateY^{0,\overline{s}_{1,k,j}+\widetilde{s}_{1,k,l}}\right)_{0\leq
k\leq N/15-1 \atop 0\leq j\leq 7}$ and
$\left(RotateY^{0,\overline{s}_{2,k,j}+\widetilde{s}_{2,k,l}}\right)_{0\leq
k\leq N/15-1 \atop 0\leq j\leq 7}$. According to how
$\overline{s}_{1,k,j}$ and $\overline{s}_{2,k,j}$ are determined in
the encryption process, we can get
$\mathbb{S}_{1,k}=\{\overline{s}_{1,k,j}\dotplus
\widetilde{s}_{1,k,l}\}_{j=0}^7\subseteq\mathbb{S}_1=\{\alpha_1\dotplus\widetilde{s}_{1,k,l},
8-\alpha_1\dotplus\widetilde{s}_{1,k,l},
\alpha_1+\beta_1\dotplus\widetilde{s}_{1,k,l},8-(\alpha_1+\beta_1)\dotplus\widetilde{s}_{1,k,l}\}$
and $\mathbb{S}_{2,k}=\{\overline{s}_{2,k,j}\dotplus
\widetilde{s}_{2,k,l}\}_{j=0}^7\subseteq\mathbb{S}_2=\{\alpha_2\dotplus\widetilde{s}_{2,k,l},
8-\alpha_2\dotplus\widetilde{s}_{2,k,l},
\alpha_2+\beta_2\dotplus\widetilde{s}_{2,k,l},8-(\alpha_2+\beta_2)\dotplus\widetilde{s}_{2,k,l}\}$.
Comparing $\mathbb{S}_1$, $\mathbb{S}_2$ with $\mathbb{R}_1$,
$\mathbb{R}_2$, we may be able to determine the values of
$\widetilde{s}_{1,k,l}$ and $\widetilde{s}_{2,k,l}$. There are four
different cases:
\begin{itemize}
\item
$\mathbb{S}_{m,k}\subset\mathbb{S}_m$: If $\mathbb{S}_{m,k}$ does
not contain all elements in $\mathbb{S}_m$, it is generally
impossible to uniquely determine $\widetilde{s}_{m,k,l}$. From
Proposition~\ref{proposition:ProbX}, the occurrence probability of
this case is $2/2^8=1/2^7$.

\item
$\mathbb{S}_{m,k}=\mathbb{S}_m$ and $\mathbb{R}_m=\{2,6\}$: When
$\widetilde{s}_{m,k,l}\in\{1,2,3,5,6,7\}$, its value can be uniquely
determined. When $\widetilde{s}_{m,k,l}=0$ or 4, it is impossible to
distinguish one value from the other.

\item
$\mathbb{S}_{m,k}=\mathbb{S}_m$ and
$\mathbb{R}_m=\{1,7\},\{3,5\},\{4,1,7\},\{4,2,6\},\{4,3,5\},\{1,2,6,7\}$
or $\{2,3,5,6\}$: The value of $\widetilde{s}_{m,k,l}$ can always be
uniquely determine.

\item
$\mathbb{S}_{m,k}=\mathbb{S}_m$ and $\mathbb{R}_m=\{1,3,5,7\}$: The
value of $\widetilde{s}_{m,k,l}$ can never be uniquely determined.
One can only determine which of the following two sets
$\widetilde{s}_{m,k,l}$ belongs to: $\{0,2,4,6\}$ and $\{1,3,5,7\}$.
\end{itemize}
Assuming the value of $\widetilde{s}_{m,k,l}$ distributes uniformly
over $\{0,\ldots,7\}$, the probability that each
$\widetilde{s}_{m,k,l}$ cannot be uniquely determined is
$1/2^7+(1-1/2^7)((1/21)(2/8)+4/21)\approx 0.2086$. We may choose
more different values of $l$ in Sec.~\ref{section:EBRS:VBR} to
decrease this probability, but the probability has a lower bound
$1/2^7+(1-1/2^7)(4/21)\approx 0.1968$. We can see this probability
is always not sufficiently small, so we will not be able to uniquely
determine the value of $\widetilde{s}_{1,k,l}$ or that of
$\widetilde{s}_{2,k,l}$ for quite a lot of blocks.

\subsubsection{Determining the secret bits controlling the 9th to 35th byte-swapping operations}

In case $\widetilde{s}_{1,k,l}$ and $\widetilde{s}_{2,k,l}$ can be
uniquely determined, we will be able to uniquely recover the 9th to
35th byte-swapping operations, i.e., we can determine the values of
$(\widetilde{s}_{1,k,i})_{i=0}^7$ and
$(\widetilde{s}_{2,k,i})_{i=0}^7$. Note
$(\widetilde{s}_{1,k,i})_{i=0}^7$ and
$(\widetilde{s}_{2,k,i})_{i=0}^7$ actually define two permutation
maps over $\{0,\ldots,7\}$. Observing the 9th to 35th byte-swapping
operations in \textit{Step b)}, one can notice that the permutation
maps has a strong pattern: 12 byte-swapping operations for the first
half-block and the other 12 ones for the second half-block, and each
group of 12 byte-swapping operations can be divided into three
phases. For the 12 byte-swapping operations performed on the first
half-block, the three phases are as follows:
\begin{itemize}
\item
\textit{Phase 1}: $(i,j,l)=(0,4,12),(1,5,13),(2,6,14),(3,5,15)$;

\item
\textit{Phase 2}: $(i,j,l)=(0,2,20), (1,3,21), (4,6,22), (5,7,23)$;

\item
\textit{Phase 3}: $(i,j,l)=(0,1,28), (2,3,29), (4,5,30), (6,7,31)$.
\end{itemize}
Apparently, Phase 1 swaps the bytes in the two 4-byte quarter-block
of the first 8-byte half-block, and Phases 2 and 3 only permute the
bytes with each 4-byte quarter-block. Then, for $i=0,1,2,3$, we can
check in which quarter-block $\overline{f^{(*16)}}(k,i)$ belongs to
after the byte-swapping operations. In other words, we check if
$\widetilde{s}_{1,k,i}\in\{0,1,2,3\}$ or $\{4,5,6,7\}$, which
corresponds to $b(129k+12+i)=0$ and 1, respectively. This allows us
to completely determine $(b(129k+12+i))_{i=0}^3$, i.e., to break
Phase 1. Then, we can derive a new permutation map represented by
$(\widetilde{s}_{1,k,i}^*)_{i=0}^7$, which consists of only Phases 2
and 3. Then, according to the byte swapping operations involved in
Phases 2 and 3, we can derive the following rule to break the 4
controlling bits involved in Phase 2:
\begin{itemize}
\item
when $i=0,1$:
$b(129k+20+i)=\begin{cases}%
0, & \widetilde{s}_{1,k,i}^*\in\{0,1\},\\
1, & \widetilde{s}_{1,k,i}^*\in\{2,3\};
\end{cases}$

\item
when $i=2,3$:
$b(129k+20+i)=\begin{cases}%
0, & \widetilde{s}_{1,k,i}^*\in\{4,5\},\\
1, & \widetilde{s}_{1,k,i}^*\in\{6,7\}.
\end{cases}$
\end{itemize}
After breaking both Phases 1 and 2, we can immediately break the 4
controlling bits $(b(129k+28+i))_{i=0}^3$ involved in Phase 3. Now,
we completely break all the 12 controlling bits involved in the
byte-swapping operations performed on the first half-block. The same
process can be applied to the second half-block, and 12 controlling
bits can be uniquely determined. As a whole, we will be able to
break all the 24 controlling bits $(b(129k+i))_{i=12}^{35}$.

\subsubsection{Determining the secret bits controlling value masking}

In case $\widetilde{s}_{1,k,l}$ and $\widetilde{s}_{2,k,l}$ can be
uniquely determined as described in Sec.~\ref{section:BreakingVBR2},
we will be able to determine $(Seed^*(k, j))_{j=0}^{15}$, or
equivalently, $(Seed(k,j))_{j=0}^8$. This allows us to obtain
$\{Seed(k,j)\}_{j=0}^8\subseteq\{Seed1(k),\overline{Seed1}(k),Seed2(k),\overline{Seed2}(k)\}$.
To break the controlling bits, we need to recover $Seed1(k)$ and
$Seed2(k)$, which are calculated from $(b(129k+i))_{i=0}^{63}$ and
$(b(129k+64+i))_{i=0}^{63}$, respectively. Note that we can always
break $(b(129k+i))_{i=0}^{35}$ if $\widetilde{s}_{1,k,l}$ and
$\widetilde{s}_{2,k,l}$ are uniquely determined. This means that we
can break the $36/4=9$ least significant bits (LSBs) of
$Seed1(k)$, since each bit of $Seed1(k)$ is determined by
four controlling bits. Then, if the nine LSBs of $Seed1(k)$ are not
all equal to those of $Seed2(k)$ or those of $\overline{Seed2}(k)$,
we can uniquely determine $Seed1(k)$ and then $\overline{Seed1}(k)$.
Assuming $Seed1(k)$ and $Seed2(k)$ are independent of each other and
each bit distributes uniformly over $\{0,1\}$, the probability that
$Seed1(k)$ cannot be uniquely determined is $2/2^9=1/2^8$. In case
$Seed1(k)$ is uniquely determined, we have the following results:
\begin{itemize}
\item
when $Seed(k,j)\in\{Seed1(k),\overline{Seed1}(k)\}$:

$b(129k+36+2j)=1$; $b(129k+37+2j)=\begin{cases}0,&
Seed(k,j)=\overline{Seed1}(k),\\1,& Seed(k,j)=Seed1(k);\end{cases}$

\item
when $Seed(k,j)\in\{Seed2(k),\overline{Seed2}(k)\}$:

$b(129k+36+2j)=1$; $b(129k+37+2j)=\begin{cases}0,&
Seed(k,j)=\overline{Seed2}(k),\\1,& Seed(k,j)=Seed2(k).\end{cases}$

Note that in this case, $Seed2(k)$ has to be guessed from the set
$\{Seed2(k),\overline{Seed2}(k)\}$.
\end{itemize}

\subsubsection{Determining the secret bits controlling horizontal/vertical bit rotations}

In case $\widetilde{s}_{1,k,l}$ and $\widetilde{s}_{2,k,l}$ can be
uniquely determined as described in Sec.~\ref{section:BreakingVBR2},
we will be able to uniquely determine the horizontal and vertical
bit rotations exerted on $\mymatrix{M}_1$,
$\widetilde{\mymatrix{M}}_1$, $\mymatrix{M}_2$ and
$\widetilde{\mymatrix{M}}_2$. Depending on how well the values of
$\alpha_1,\beta_1,\alpha_2,\beta_2$ are determined in
Sec.~\ref{section:BreakingAlphaBeta12}, some information about the
controlling bits involved in the bit rotations may be obtained,
although it is always impossible to uniquely determine the value of
any controlling bit involved. Since the determination process of the
controlling bits are similar for $\mymatrix{M}_1$,
$\widetilde{\mymatrix{M}}_1$, $\mymatrix{M}_2$ and
$\widetilde{\mymatrix{M}}_2$, here we consider only the case of
$\mymatrix{M}_1$ (i.e., horizontal bit rotations exerted on the
first half-block) to simplify the discussion. For this case, we get
$(\overline{r}_{1,k,i})_{i=0}^7$ by substituting
$\widetilde{r}_{1,k,l}$ into $(\overline{r}_{1,k,(i\dotplus
\widetilde{r}_{1,k,l})})_{i=0}^7$. In Step d),
$\overline{r}_{1,k,i}$ is determined by two controlling bits as
follows:
\[
\overline{r}_{1,k,i}=\begin{cases}
\alpha_1, & (b(129k+65+2i),b(129k+66+2i))=(0,0),\\
\alpha_1+\beta_1, & (b(129k+65+2i),b(129k+66+2i))=(0,1),\\
8-\alpha_1, & (b(129k+65+2i),b(129k+66+2i))=(1,0),\\
8-(\alpha_1+\beta_1), & (b(129k+65+2i),b(129k+66+2i))=(1,1).
\end{cases}
\]
We have the following different cases.
\begin{itemize}
\item
$\mathbb{R}_1=\{1,7\},\{2,6\}$ or $\{3,5\}$: In this case,
$\alpha_1$ and $\beta_1$ can be uniquely determined, but we cannot
differentiate $\alpha_1$ from $8-(\alpha_1+\beta_1)$, and
$8-\alpha_1$ from $\alpha_1+\beta_1$. Hence, we can determine
neither $b(129k+65+2i)$ nor $b(129k+66+2i)$, but just the following:
\[
(b(129k+65+2i),b(129k+66+2i))=\begin{cases}%
(0,0)\text{ or }(1,1), & \overline{r}_{1,k,i}\in\{1,2,3\},\\
(0,1)\text{ or }(1,0), & \overline{r}_{1,k,i}\in\{5,6,7\}.
\end{cases}
\]

\item
$\mathbb{R}_1=\{4,1,7\},\{4,2,6\}$ or $\{4,3,5\}$: In this case,
$(\alpha_1,\beta_1)$ has two possible values, so
$(b(129k+65+2i),b(129k+66+2i))$ cannot be uniquely determined. What
we can get is the following:
\[
(b(129k+65+2i),b(129k+66+2i))=\begin{cases}%
(0,0)\text{ or }(1,1), & \overline{r}_{1,k,i}\in\{1,2,3\},\\
(0,1)\text{ or }(1,0), & \overline{r}_{1,k,i}\in\{5,6,7\},\\
(0,0),(0,1),(1,0)\text{ or }(1,1), & \overline{r}_{1,k,i}=4.
\end{cases}
\]

\item
$\mathbb{R}_1=\{1,2,6,7\}$: In this case, $(\alpha_1,\beta_1)$ has
four possible values $(1,1)$, $(1,5)$, $(2,5)$ or $(6,1)$, so
$(b(129k+65+2i),b(129k+66+2i))$ cannot be uniquely determined,
either. What we can get is the following:
\[
(b(129k+65+2i),b(129k+66+2i))=\begin{cases}%
(0,0)\text{ or }(1,1), & \overline{r}_{1,k,i}=1,\\
(0,1)\text{ or }(1,0), & \overline{r}_{1,k,i}=7,\\
(0,0),(0,1),(1,0)\text{ or }(1,1), & \overline{r}_{1,k,i}\in\{2,6\}.
\end{cases}
\]

\item
$\mathbb{R}_1=\{1,3,5,7\}$: In this case, $(\alpha_1,\beta_1)$ has
four possible values $(1,2)$, $(1,4)$, $(3,4)$ or $(5,2)$, so
$(b(129k+65+2i),b(129k+66+2i))$ cannot be uniquely determined,
either. What we can get is the following:
\[
(b(129k+65+2i),b(129k+66+2i))=\begin{cases}%
(0,0)\text{ or }(1,1), & \overline{r}_{1,k,i}=1,\\
(0,1)\text{ or }(1,0), & \overline{r}_{1,k,i}=7,\\
(0,0),(0,1),(1,0)\text{ or }(1,1), & \overline{r}_{1,k,i}\in\{3,5\}.
\end{cases}
\]

\item
$\mathbb{R}_1=\{2,3,5,6\}$: In this case, $(\alpha_1,\beta_1)$ has
four possible values $(2,1)$, $(2,3)$, $(3,3)$ or $(5,1)$, so
$(b(129k+65+2i),b(129k+66+2i))$ cannot be uniquely determined,
either. What we can get is the following:
\[
(b(129k+65+2i),b(129k+66+2i))=\begin{cases}%
(0,0)\text{ or }(1,1), & \overline{r}_{1,k,i}=2,\\
(0,1)\text{ or }(1,0), & \overline{r}_{1,k,i}=6,\\
(0,0),(0,1),(1,0)\text{ or }(1,1), & \overline{r}_{1,k,i}\in\{3,5\}.
\end{cases}
\]
\end{itemize}

\subsubsection{Summary}

As a brief summary, based on the equivalent key obtained in the
differential attack, we can further determine
$\mathbb{R}_1=\{\alpha_1,8-\alpha_1,\alpha_1+\beta_1,8-(\alpha_1+\beta_1)\}$
and
$\mathbb{R}_2=\{\alpha_2,8-\alpha_2,\alpha_2+\beta_2,8-(\alpha_2+\beta_2)\}$
with a very high probability $1-1/2^{8N/15-1}$. Then, we may be able
to uniquely determine the value of $(\alpha_m,\beta_m)$ ($m=1,2$)
with probability $3/21=1/7$, or narrow down the number of possible
values to 2 (with probability $6/21=2/7$) or to 4 (with probability
$12/21=4/7$). Based on $\mathbb{R}_m$ ($m=1,2$), we may be able to
recover $\widetilde{s}_{m,k,l}$ with probability $\geq
1-0.1968\approx 0.8032$. In case $\widetilde{s}_{1,k,l}$ and
$\widetilde{s}_{2,k,l}$ are uniquely determined, we have the
following results:
\begin{itemize}
\item
Controlling bits $(b(129k+i))_{i=12}^{35}$ can always be uniquely
determined.

\item
In case the value of $Seed1(k)$ can be recovered, which happens with
probability $1-1/2^8$, the controlling bits
$(b(129k+36+2j))_{j=0}^7$ can always be uniquely determined, but
$(b(129k+37+2j))_{j=0}^7$ can be uniquely determined only when
$Seed(k,j)\in\{Seed1(k),\overline{Seed1}(k)\}$.

\item
None of the controlling bits involved in the bit rotations can be
uniquely determined, but we may be able to narrow down the number of
possible values of the two controlling bits determining
each bit-rotation operation from 4 to 2 in some cases.
\end{itemize}

\color{black}

\subsection{Experimental results}

\begin{figure}[!htb]
\centering
\begin{minipage}{\figwidth}
\centering
\includegraphics[width=\textwidth]{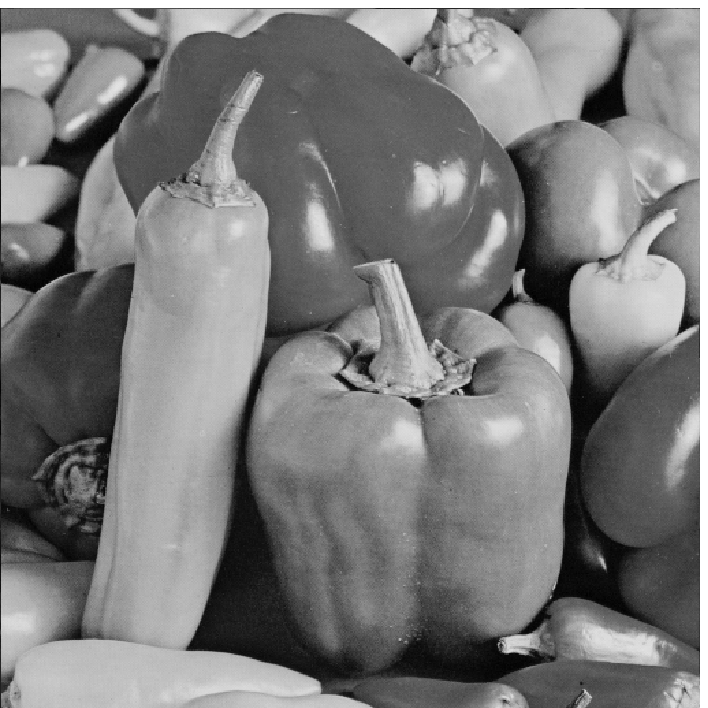}
a)
\end{minipage}
\begin{minipage}{\figwidth}
\centering
\includegraphics[width=\textwidth]{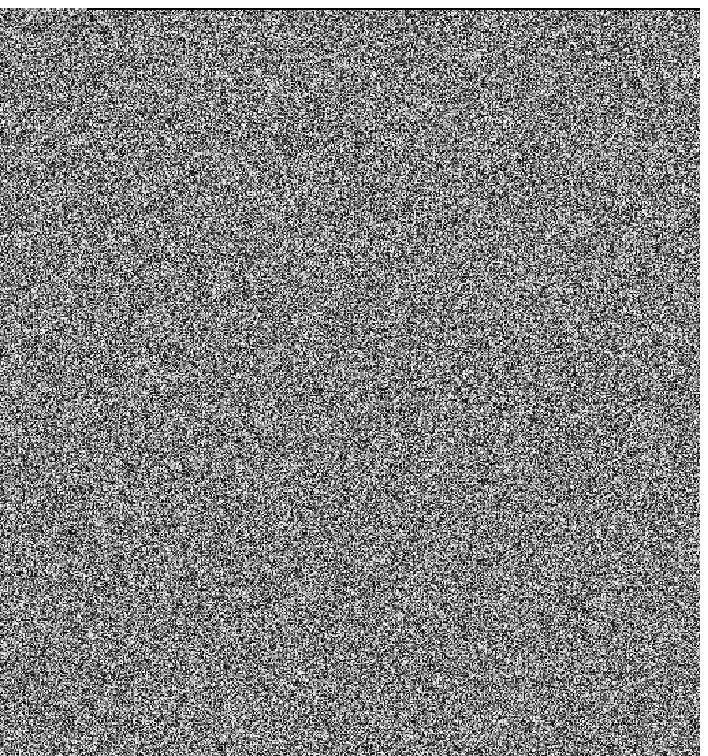}
b)
\end{minipage}
\caption{The plain-image ``Peppers'' and the corresponding
cipher-image: a) the plain-image; b) the cipher
image.}\label{figure:encryptpeppers}
\end{figure}

\begin{figure}[!htb]
\centering
\begin{minipage}{\figwidth}
\centering
\includegraphics[width=\textwidth]{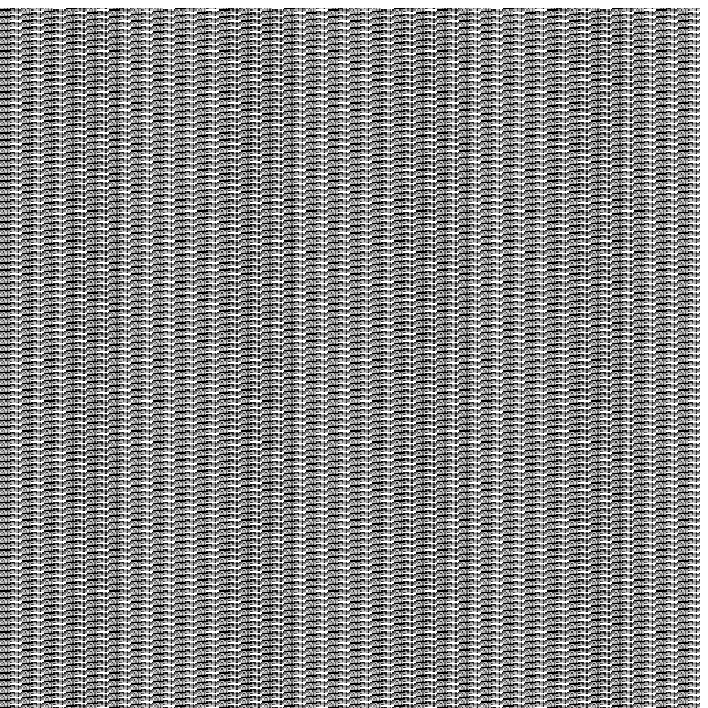}
a)
\end{minipage}
\begin{minipage}{\figwidth}
\centering
\includegraphics[width=\textwidth]{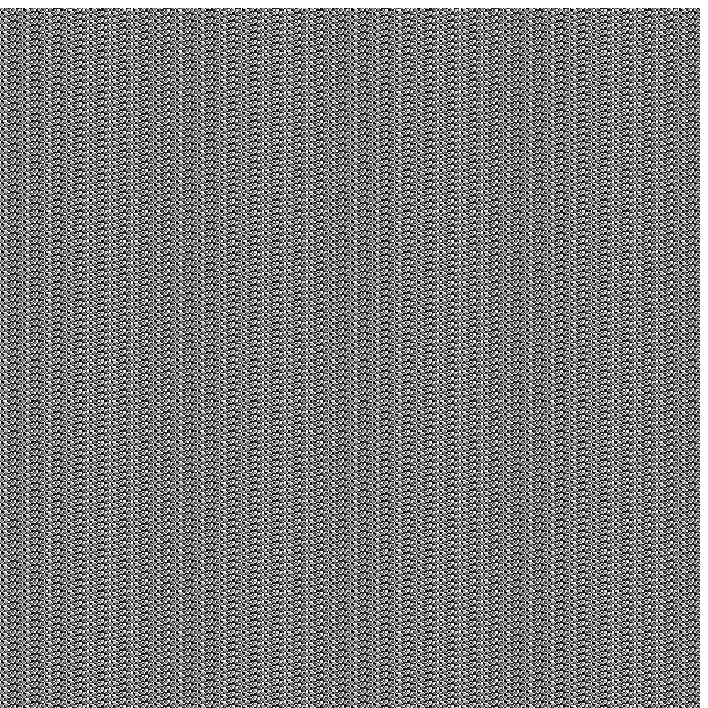}
b)
\end{minipage}
\caption{The two plaintext differentials for breaking data
expansion.}\label{figure:differentialfile1}
\end{figure}

\begin{figure}[!htb]
\centering
\begin{minipage}{\figwidth}
\centering
\includegraphics[width=\textwidth]{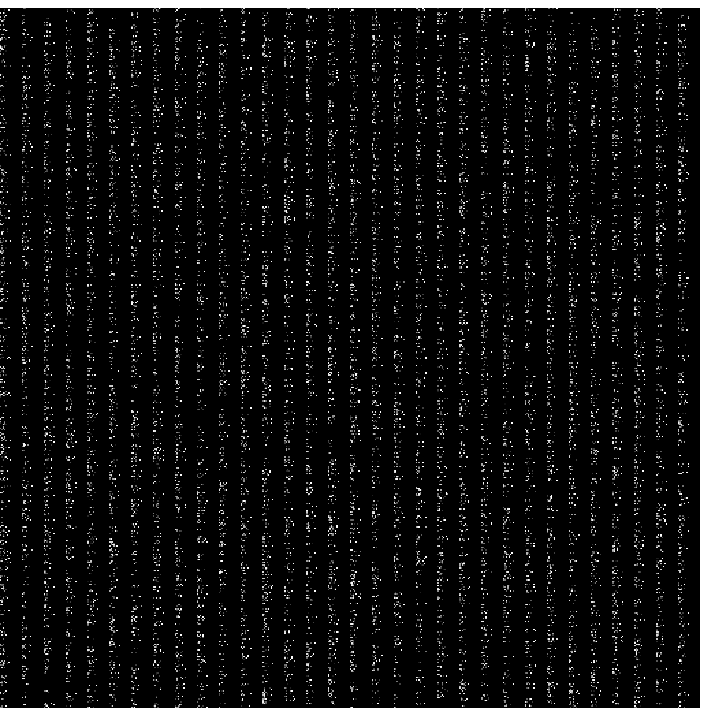}
a)
\end{minipage}
\begin{minipage}{\figwidth}
\centering
\includegraphics[width=\textwidth]{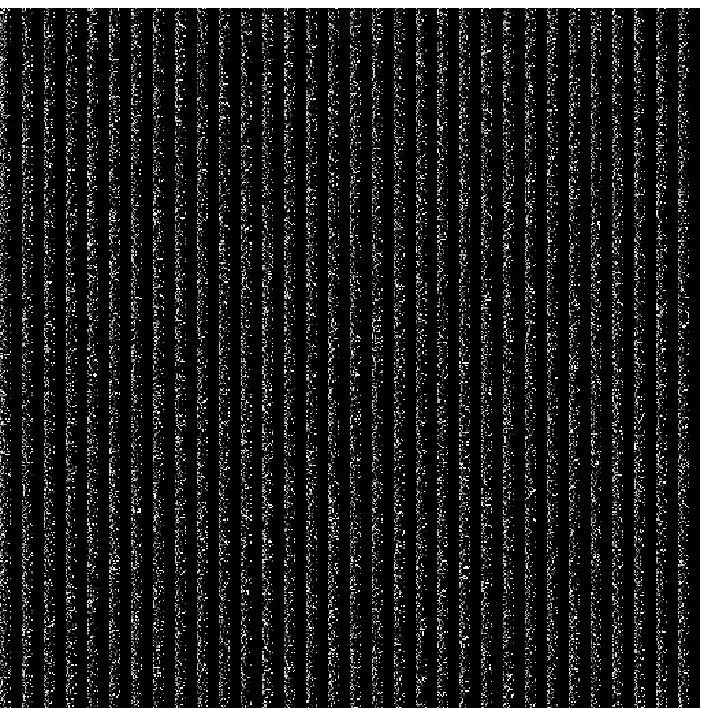}
b)
\end{minipage}
\caption{The two plaintext differentials for breaking the first
eight byte-swapping operations.}\label{figure:differentialfile2}
\end{figure}

\begin{figure}[!htb]
\centering
\begin{minipage}{\figwidth}
\centering
\includegraphics[width=\textwidth]{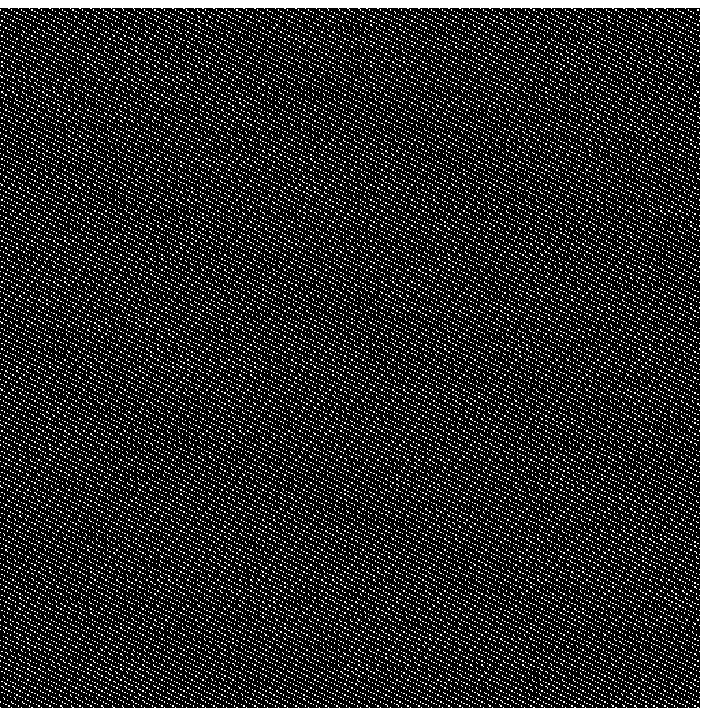}
a)
\end{minipage}
\begin{minipage}{\figwidth}
\centering
\includegraphics[width=\textwidth]{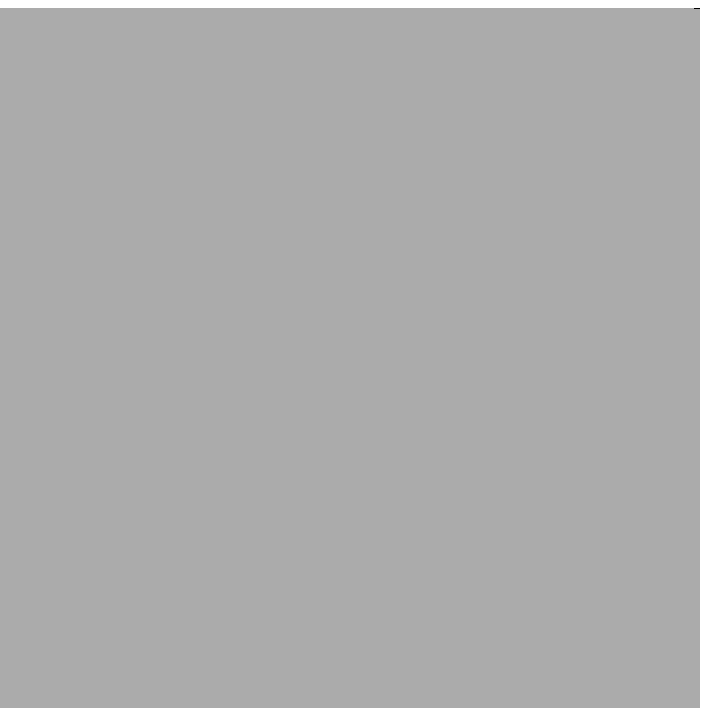}
b)
\end{minipage}
\caption{The two plaintext differentials for obtaining the vertical
and horizontal bit-operation part of the EES: a) vertical
bit-operation; b) horizontal
bit-operation.}\label{figure:differentialfile3}
\end{figure}

\begin{figure}[!htb]
\centering
\begin{minipage}{\figwidth}
\centering
\includegraphics[width=\textwidth]{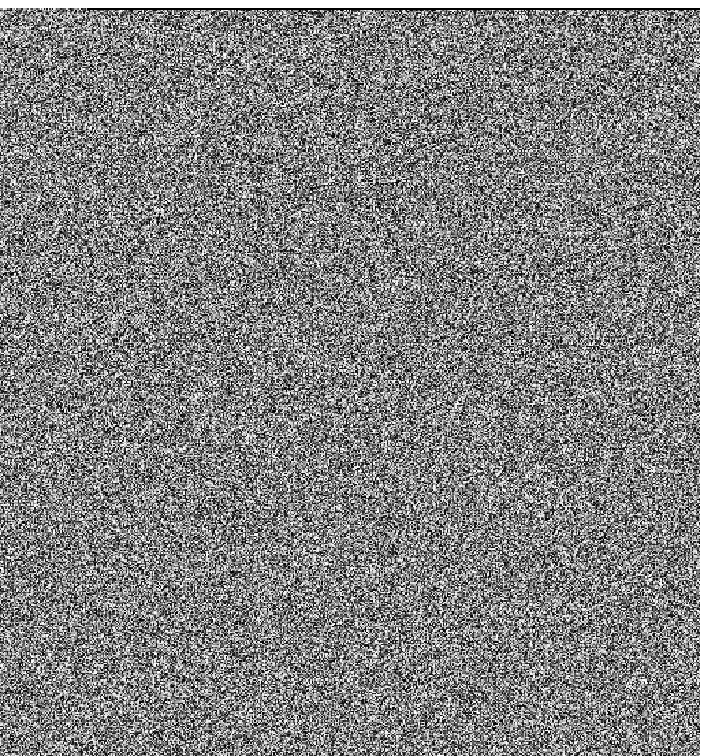}
a)
\end{minipage}
\begin{minipage}{\figwidth}
\centering
\includegraphics[width=\textwidth]{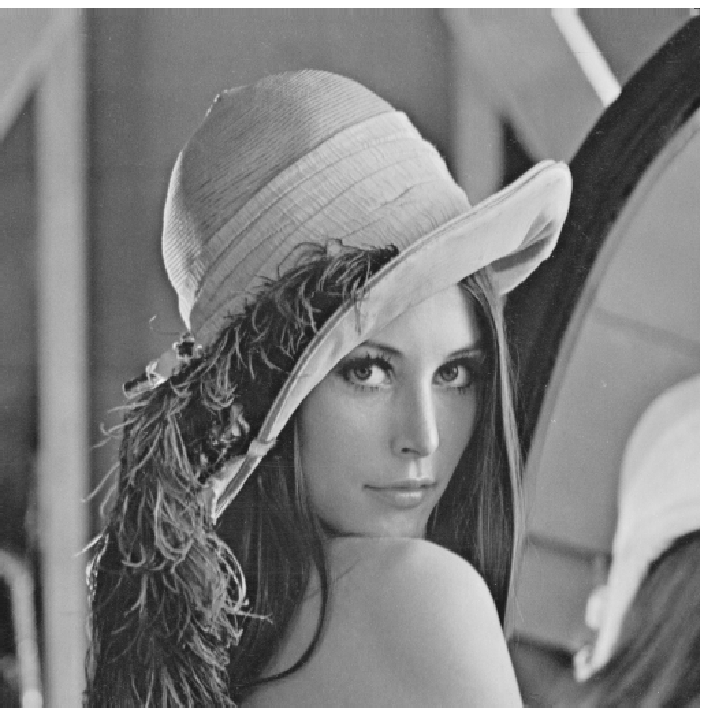}
b)
\end{minipage}
\caption{The decryption result of another cipher-image encrypted
with the same secret key: a) cipher-image; b) decrypted
plain-image.}\label{figure:decrypt}
\end{figure}

To verify the real performance of the differential attack proposed
in this paper, some experiments were carried out with the following
randomly selected secret key: $\alpha_1=2$, $\beta_1=5$,
$\alpha_2=3$, $\beta_2=4$, $Secret=20$, and $x(0)=0.251$.
Figure~\ref{figure:encryptpeppers} shows a $512\times 512$
plain-image ``Peppers'' and the corresponding cipher-image. Note
that the cipher-image is $1/16$ higher than the plain-image due to
the data expansion. This plain-image is used as one of the chosen
plaintext $f_0$ to generate the required chosen plaintext
differentials. The two differentials used for breaking secret data
expansion are shown in Figs.~\ref{figure:differentialfile1}. The two
differentials used for breaking the first eight byte-swapping
operations, i.e., the secret bits $\{b(129k+i)\}_{0\leq k\leq N/15-1
\atop 0\leq k\leq 7}$, are shown in
Fig.~\ref{figure:differentialfile2}. The two differentials shown in
Fig.~\ref{figure:differentialfile3} and those two shown in
Figs.~\ref{figure:differentialfile1} were used to obtain an EES. The
recovered equivalent key (i.e., all the items shown in
Sec.~\ref{section:AttackPerformance}) was used to decrypt a
cipher-image as shown in Fig.~\ref{figure:decrypt}a). The result is
given in Fig.~\ref{figure:decrypt}b). It can be seen that the secret
plain-image was successfully recovered by the differential attack.

\section{Conclusion}

In this paper, we evaluate the security of a recently-proposed
multimedia encryption system called MCS
\cite{Yen-Guo:MCS:ISCAS2005}, and propose a differential attack to
break it with a divide-and-conquer (DAC) strategy. The
differential attack is very efficient in the sense that only seven
chosen plaintexts are needed to get an equivalent key and the
computational complexity is only $O(N)$, where $N$ is the number of
bytes in the plaintext. The real performance of the proposed attack
was also verified with experiments. Similar to some other image
encryption schemes proposed in the literature, the MCS was not
designed by following some good principles of designing such
systems. Some of these principles are discussed in
\cite{AlvarezLi:Rules:IJBC2006,Li:AttackingRCES2008}.

\bibliographystyle{IEEEtran}
\bibliography{IEEEabrv,MCS}

\end{document}